\documentclass[10pt,journal,compsoc]{IEEEtran}

\usepackage{graphicx}
\usepackage{multirow}
\usepackage{color}
\usepackage{algorithm}
\usepackage{amsmath}
\usepackage{amssymb}
\usepackage{gensymb}
\usepackage{xspace}
\usepackage{algpseudocode}
\usepackage{algorithmicx}
\usepackage{balance}
\usepackage{lastpage}
\usepackage{enumerate}

\begin{document}
\title{EOP: An Encryption-Obfuscation Solution for Protecting PCBs Against Tampering and Reverse Engineering}

\author{\IEEEauthorblockN{Zimu Guo, Xiaolin Xu, Mark M. Tehranipoor and Domenic Forte}
\IEEEauthorblockA{ECE Department, University of Florida\\
Email: zimuguo@ufl.edu; \{xiaolinxu,tehranipoor,dforte\}@ece.ufl.edu}
}

\maketitle

\begin{abstract}
PCBs are the core components for the devices ranging from the consumer electronics to military applications. Due to the accessibility of the PCBs, they are vulnerable to the attacks such as probing, eavesdropping, and reverse engineering. In this paper, a solution named EOP is proposed to migrate these threats. EOP encrypts the inter-chip communications with the stream cipher. The encryption and decryption are driven by the dedicated clock modules. These modules guarantee the stream cipher is correctly synchronized and free from tampering. Additionally, EOP also incorporates the PCB-level obfuscation for protection against reverse engineering. EOP is designated to be accomplished by utilizing the COTS components. For the validation, EOP is implemented in a Zynq SoC based system. Both the normal operation and tampering detection performance are verified. The results show that EOP can deliver the data from one chip to another without any errors. It is proved to be sensitive to any active tampering attacks.
\end{abstract}

\IEEEpeerreviewmaketitle

\section{Introduction}
\label{sec:intro}
Printed circuit boards (PCBs) are essential components for electronic systems, ranging from consumer electronics to military applications. Due to the PCBs' proprieties (e.g., large size, exposed connections, etc.), they are vulnerable to various attacks like probing, eavesdropping, hardware tampering, and reverse engineering. Tampering attacks target the exposed PCB-level connections (e.g., copper traces, component solder pads) and can be classified into two types: passive and active \cite{ghosh2015secure}. Passive tampering is implemented by monitoring sensitive data traffic between electronic components. Active tampering works by injecting malicious data/instructions into the system, modifying the PCB, etc. Furthermore, active tampering attacks can diminish the strength of the obfuscation-based protection against reverse engineering \cite{forte2017hardware}. The rapid spread of IoT devices could empower tampering attacks to induce greater damages. For instance, a tampered device with the internet accessibility can be deployed as a botnet to execute distributed denial-of-service (DDoS) attacks, steal sensitive/private data, and even allow the attacker to access the connections of the botnets. Additionally, by tampering the consumer electronics (e.g., video game consoles), illegal users can bypass vendor copyright \cite{bluraymod}.

\begin{figure}[b]
  \centering
  \includegraphics[width=\linewidth]{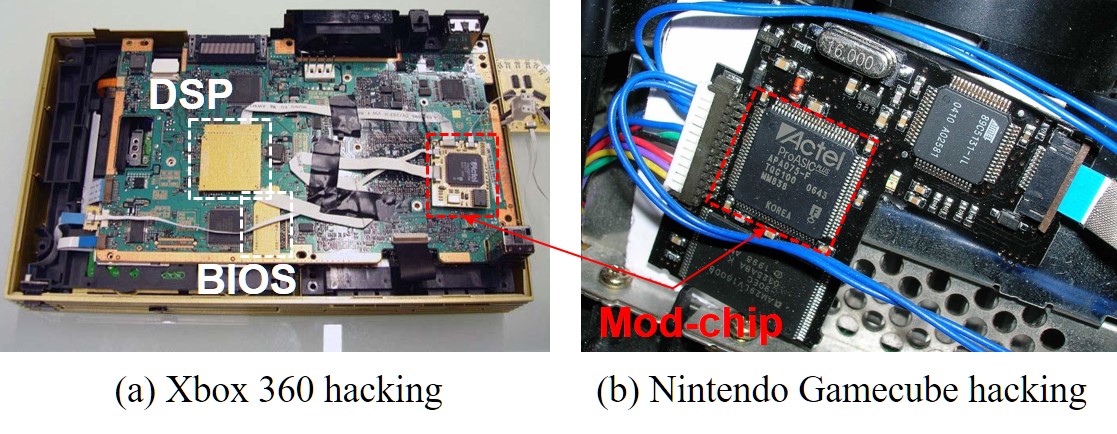}
  \caption{Modchip installed on the video game consoles (s) Xbox 306 and (b) Nintendo Gamecube. The modchip is marked in the red rectangle.}
  \label{fig:modchip example}
\end{figure}

A common implementation of the active tampering attack can be implemented by installing a modchip (modification chip). \textcolor{black}{A modchip is a small electronic device that can be used to alter or disable the artificial restrictions of computers or entertainment devices. Modchips are mainly utilized in tampering the video game consoles (as shown in Figure \ref{fig:modchip example}) and DVD or Blu-ray players \cite{bluraymod}.} Besides media devices, routers and networking equipment on route to foreign locations were reportedly intercepted and tampered by the National Security Agency (NSA) \cite{cisco_upgrade} for surveillance purposes. A load station installed a beacon firmware to monitor and redirect user data. The equipment was then resealed and sent to its destination.

In \cite{guo2015investigation}, a framework was developed to hide critical inter-chip traces within the middle layers of a multi-layer PCB. The traces chosen are ones connected to chips with the ball grid array (BGA) package. The pins of this type of package are located at the bottom of the chip. While combining the middle-layer routing and BGA package, both the passive and active tampering attacks can be prevented by reducing the physical accessibility to the traces. However, these design requirements limit the applicability of this approach. To accomplish passive tampering detection, a board-level framework, named MPA, was introduced in \cite{guo2017mpa}. This framework provides an accurate measurement of the trace impedance changes. By applying this approach, even the tiny changes induced by the passive tampering can be detected. The major drawbacks of this approach are (a) a dedicated detection phase is required prior to the normal operation, and (b) the genuine device should be enrolled before sold in the market. Thus, this approach is not appropriate for the high-volume applications (e.g., IoT).

Another PCB vulnerability to be seriously concerned about is reverse engineering. The techniques for accomplishing this attack can be classified as destructive and non-destructive \cite{forte2017hardware}. While comparing with the chip-level reverse engineering, it is much easier to apply such attacks at the PCB level. The reverse engineering attacks cause the disclosure of the PCB design, enabling copies to enter the market as well as providing information for other physical attacks.

Previous physical security requirements of known security standards (Federal Information Processing Standard (FIPS 140-2) \cite{caddy2011fips}, and Payment Card Industry Hardware Security Module (PCI HSM) \cite{mavrovouniotis2014hardware}) do not specifically address the tampering and reverse engineering attacks. Thus, they cannot provide practical guidance for design verification within this scenario.

In this paper, we propose an encryption-obfuscation based protection to provide the PCB-level assurance (EOP). EOP encrypts the real-time communication between critical inter-chip connections. More specifically, the encryption is realized by applying XOR logic between the plaintext message and the one-time-pad (OTP). \textcolor{black}{In order to select a proper cryptography system which is suitable for the real-time applications, we investigated various cryptography techniques, such as the block and stream ciphers.} A hardware-oriented stream cipher is implemented as the OTP. It is synchronized by a control clock, which is generated and verified by the dedicated modules. \textcolor{black}{The EOP is designated to incorporate with the systems with the commercial-off-the-shelf (COTS) components.} The major contributions of this work are:

\begin{itemize}
  \item The EOP framework is proposed to protect the critical data paths of PCBs from tampering attacks. This framework can be combined with other existing approaches, such as obfuscation, to prevent the board-level reverse engineering.
  \item Based on the tampering protection targets and goals, we investigate the appropriateness and security of the available ciphers (e.g., block ciphers and stream ciphers). A non-linear stream cipher is implemented for EOP.
  \item A secure synchronization strategy of the stream cipher is proposed. \textcolor{black}{This strategy consists of a key management/initialization phase.} By monitoring the plaintext and encrypted messages, this strategy generates and verifies the stream cipher synchronization signals.
  \item EOP is implemented on the Xilinx SoCs, and its performance against various tampering attacks is validated.
\end{itemize}

The rest of the paper is organized as follows. Section \ref{sec:background} provides the background and comparison of various types of ciphers. Moreover, the best cipher implementation candidate to protect the PCB-level secret is introduced with details. In Section \ref{sec:method}, security requirements are presented as the protection design guidelines. Following these guidelines, the details of EOP are given. \textcolor{black}{The design of the cryptography and other control models are also presented in this section.} Besides the framework, its robustness against various attacks is also discussed in Section \ref{sec:attacks analysis}. The experimental setup and validation results are presented in Section \ref{sec:resutls}. Finally, we conclude in Section \ref{sec:conclusion}.

\section{Background}
\label{sec:background}
In this section, we first review several ciphers and compare their applicability for real-time systems. Next, a hardware-oriented cipher, which will be implemented in EOP, is introduced. Finally, a brief introduction of the obfuscation-based PCB anti-reverse engineering scheme (referred as PCB obfuscation scheme for the rest of the paper) is provided.

\subsection{Block Cipher vs. Stream Cipher}
\label{sec:block vs stream}
In cryptography, cipher designs can be classified into two categories depending on how the plaintext is partitioned \cite{stinson2005cryptography}. In a \textbf{block cipher}, the plaintext is divided into relatively large (e.g., 128 bit) blocks which are encoded separately. The encoding of each block depends on at most one of the previous blocks. Most block cipher algorithms are classified as iterated block ciphers where fixed-size blocks of plaintext are transformed into identical size blocks of ciphertext, via the repeated application of an invertible transformation. Data Encryption Standard (DES) and Advanced Encryption Standard (AES) are well-known block ciphers. DES has a fixed block size of 64 bits, while the block size for AES is 128 bits. For each block, the same key is applied.

Different from the large block size of the block cipher, \textbf{stream ciphers} operate on plaintext in a bitwise fashion \cite{cusick2004stream}. A stream cipher is a symmetric key cipher where the plaintext digits are combined with a pseudorandom cipher stream (keystream). In a stream cipher, each plaintext digit is encrypted one at a time with the corresponding digit of the keystream. This digit is typically a 1-bit data, and the encryption operation is usually based on exclusive-or (XOR). The pseudorandom keystream is typically generated serially from a random seed using digital shift registers, such as a linear feedback shift registers (LFSR). Other stream ciphers can be constructed by operating a block cipher in stream mode \cite{rueppel2012analysis}.

Both block cipher and stream cipher have their advantages and limitations, a comparison is provided in Table \ref{tab:cipher compare}. For block ciphers, information from one plaintext is diffused into several ciphertext blocks. It is difficult to insert symbols without being detected. This modification can be detected with a separate message authentication code such as CBC-MAC for each block. One major limitation of the block cipher is that an entire block must be accumulated before the encryption/decryption begins. Moreover, the block ciphers are slow due to multiple iterations for both the encryption and decryption. For the stream ciphers, the algorithm is linear in time, and the error in one symbol will not affect its subsequent symbols. However, the diffusion is low since all information of the plaintext is contained in a single ciphertext symbol. Moreover, the algorithm of the stream cipher is not designed to detect the illegal symbol insertions.

\begin{table}[t]
\centering
\caption{Comparison of the block cipher and stream cipher}
\label{tab:cipher compare}
\renewcommand{\arraystretch}{1.3}
\begin{tabular}{|c|c|c|}
\hline
                        & Block cipher & Stream cipher \\ \hline
Diffusion               & High         & Low           \\ \hline
Insertion detection     & Yes          & No            \\ \hline
Speed of transformation & Slow         & Fast          \\ \hline
Error propagation       & High         & Low           \\ \hline
\end{tabular}
\end{table}

Considering our application scenario, a primary requirement for encrypting the chip-to-chip communication is the real-time capability. For instance, assume the sender chip transmits a high-frequency control signal to the receiver chip. If a block cipher (e.g., DES) is implemented, the encryption and decryption processes introduce a significant delay to this signal transmission. This delay might be an issue for systems that have strict requirements on speed. Additionally, since the block cipher has a fixed (e.g., 64-bit) block size, the efficiency is low if the width of this control signal is less than the block size. For the stream ciphers (e.g., LFSR), a new keystream bit can be generated in each clock cycle. As a result, the stream cipher introduces negligible delay during the encryption and decryption. Besides the delay issue, the encryption and decryption of the block ciphers consist of multiple rounds. These procedures consume more resources than the stream ciphers. Thus, stream ciphers are more appropriate for encrypting and/or obfuscating the real-time communication between different chips. Hence, in this paper, we adopt a stream cipher. Note that since they are vulnerable to tampering/insertions and reverse engineering, we must introduce additional mechanisms into EOP (see Sections III-C).

Based on the mechanisms of updating the internal states, stream ciphers can be classified as linear and non-linear stream ciphers, respectively. A linear stream cipher updates the next state by applying a linear function of its previous states. A widely-implemented example is LFSR. However, the LFSR is usually considered as insecure due to the low complexity, i.e., a small fragment of the keystream can be used to deduce the entire sequence. A trivial attack example is provided in Section \ref{sec:partial encryption} against the linear stream cipher. To increase the complexity of stream ciphers, non-linear mechanisms have been developed. In the following section, a non-linear cipher is introduced and implemented in EOP.

\subsection{Trivium Cipher}
\label{sec:trivium}
Trivium is a synchronous stream cipher designed to provide a flexible trade-off between the operation speed, overhead of hardware and software implementation. It was submitted to the Profile II (hardware) of the eSTREAM competition which is organized by the EU ECRYPT network \cite{eSTREAM}. This cipher belongs to the non-linear stream cipher category and is inspired by the block cipher in stream-mode. It is initialized by loading an 80-bit key and an 80-bit initialization vector (IV) into the 288-bit initial state. An output bit can be generated after each cycle. The operation of the Trivium cipher is described in Algorithm \ref{alg:Trivium}. Note that here, and in this algorithm, the $+$ and $\cdot$ operations stand for addition and multiplication over GF(2) (i.e., XOR and AND). The output bit ($z_i$) of the cipher is generated according to the internal state ($s_1,\dots,s_{288}$). The internal state is self-updated according to its previous value. In this algorithm, the notations $t_1$, $t_2$, and $t_3$, indicate the temporary values. The width of the keystream can be easily extended up to 64-bits without increasing the number of flip-flops. Other than the flip-flops, the number of logic gates (e.g., AND and XOR gates) increases linearly with the key width. The estimated gate counts are provided in Table \ref{tab:Trivium overhead}. The Trivium cipher is proven to be secure against various attacks, such as guess and determination attacks, algebraic attacks, resynchronization attacks, etc. \cite{de2006stream}.

\begin{table}[b]
\centering
\caption{Estimated gate counts of 1-bit to 64-bit hardware implementations \cite{de2005trivium}}
\label{tab:Trivium overhead}
\renewcommand{\arraystretch}{1.3}
\begin{tabular}{|c|c|c|c|c|c|}
\hline
Key steam width & 1-bit & 8-bit & 16-bit & 32-bit & 64-bit \\ \hline
Flip-flops & 288   & 288   & 288    & 288    & 288    \\ \hline
AND gates  & 3     & 24    & 48     & 96     & 192    \\ \hline
XOR gates  & 11    & 88    & 176    & 354    & 704    \\ \hline
\end{tabular}
\end{table}

\begin{algorithm}[t]
\caption{Trivium key stream generation \cite{de2005trivium}}\label{alg:Trivium}
\begin{algorithmic}[1]
\For{$i=1$ to $K$}
    \State $t_1  \gets s_{66} +  s_{93}$
    \State $t_1 \gets s_{162} + s_{177}$
    \State $t_1 \gets s_{243} + s_{288}$
    \State \textbf{$z_1$} $\gets t_1 + t_2 + t_3$
    \State $t_1 \gets t_1 + s_{91} \cdot s_{92} + s_{171}$
    \State $t_1 \gets t_1 + s_{91} \cdot s_{92} + s_{171}$
    \State $t_1 \gets t_1 + s_{91} \cdot s_{92} + s_{171}$
    \State $(s_1,s_2,\dots,s_{93}) \gets (t_3,s_1,\dots,s_{92})$
    \State $(s_{94},s_{95},\dots,s_{177}) \gets (t_1,s_{94},\dots,s_{176})$
    \State $(s_{178},s_{279},\dots,s_{288}) \gets (t_2,s_{178},\dots,s_{287})$
\EndFor
\end{algorithmic}
\end{algorithm}

\subsection{PCB Obfuscation}
\label{sec:obfuscation}
The obfuscation-based solutions have been developed to prevent reverse engineering on both chip and board levels \cite{forte2017hardware}. In the board-level technique described in \cite{guo2015investigation}, a permutation block is inserted to shuffle the original inter-chip connections. This permutation block behaves like a router to navigate the signal from one chip to another. A key is applied to the permutation block to determine the input-output relationship. Only if a correct key is loaded will the signals be navigated to the correct path. In this case, the system is referred as operating in the functional mode. Otherwise, the system works in the obfuscated mode.

In the functional mode, the permutation block continually connects its inputs to the designated outputs. Thus, an attacker can inject a signal with a distinguishable frequency or pattern on one of the inputs of a permutation block in the functional mode. By monitoring all the outputs for this injected signal, the attacker could discover the correct input-output relationship. Applying this attack on the PCB is straightforward since the connections among chips are usually exposed and easy to tamper. Certain restrictions were proposed to restrict this attack: (i) the permutation block and at least one chip connected with it should be of BGA package; (ii) the internal connections between these two BGA chips should be routed within the middle layers of the PCB \cite{guo2015investigation}. By incorporating a stream cipher, we shall avoid such restrictions.

\section{Proposed Methodology}
\label{sec:method}
In this section, we propose several security requirements for the PCB-level protection. Next, the general requirements for implementing EOP are provided in Section \ref{sec:implementation overview}. According to these requirements, different application scenarios are presented. Finally, the major building blocks are elaborated in Section \ref{sec:encryption modules} and \ref{sec:control clock}.

\subsection{Security Requirements}
\label{sec:sr}
Before presenting EOP, the security requirements (SR) are discussed. These requirements aim to counter the hardware vulnerabilities of the digital systems, especially consumer electronics. The security requirements are listed as follows:
\begin{itemize}
  \item \textbf{SR-1}: Passive tampering protection. The device shall prevent the disclosure of sensitive information by external monitoring.
  \item \textbf{SR-2}: Active tampering protection. The device shall monitor or prevent unauthorized changes to the inter-chip signals.
  \item \textbf{SR-3}: Run-time tampering protection. The device shall constantly guarantee \textbf{SR-1} and \textbf{SR-2} during any operation stages once the device is on.
  \item \textbf{SR-4}: Reverse engineering prevention. The PCB design shall be protected from any physical reverse engineering whether power is on or off.
\end{itemize}

\textbf{SR-1} and \textbf{SR-2} define the capability of a mechanism against tampering attack. This capability can be accomplished by either preventing or detecting tampering activities. \textbf{SR-3} defines the tampering protection capability during runtime. \textbf{SR-4} ensures that the system cannot be copied by applying the reverse engineering attack.


\subsection{Implementation Overview}
\label{sec:implementation overview}
\begin{figure}[t]
  \centering
  \includegraphics[width=0.85\linewidth]{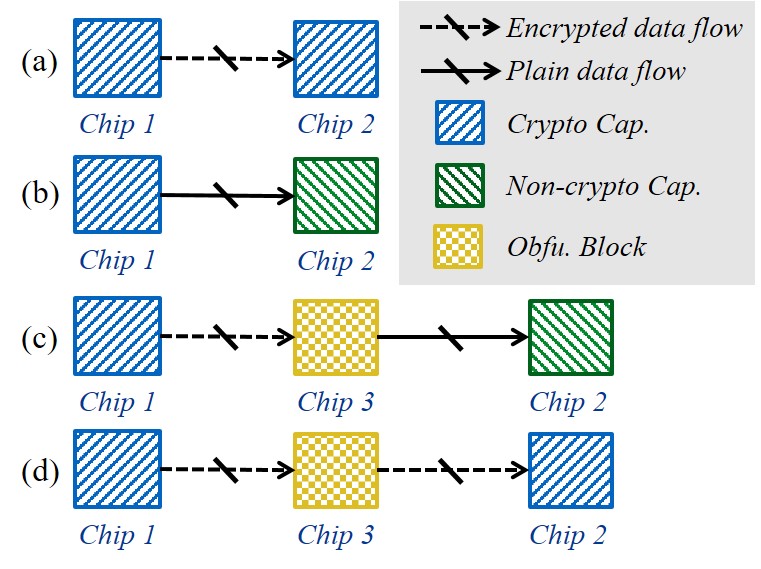}
  \caption{General requirements for implementing EOP and application scenarios: (a) fully applicable, (b) conditionally applicable, (c) obfuscation simplification, and (d) encryption-obfuscation combination.}
  \label{fig:application_scenario}
\end{figure}

The requirements for implementing the framework and application scenarios are shown in Figure \ref{fig:application_scenario}. We assume that the connections between chip 1 and chip 2 are potential targets of the tampering attack. Since EOP relies on the stream cipher, certain cryptographic capabilities are mandatory (details are discussed in Section \ref{sec:encryption modules}). According to this figure, EOP can be determined as: (a) \textbf{fully applicable}, (b) \textbf{conditionally applicable}, (c) \textbf{obfuscation simplification} and (d) \textbf{encryption-obfuscation combination}.

\begin{figure}[t]
  \centering
  \includegraphics[width=\linewidth]{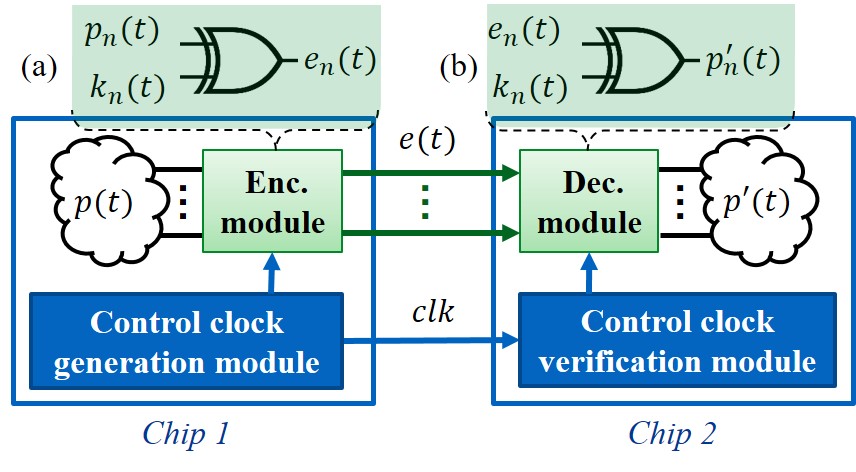}
  \caption{Block diagram of EOP. The internal logic of the encryption module (Enc. module) is presented in (a), and the internal logic of the decryption module (Dec. module) is presented in (b). The notations $p$, $e$, and $k$ stands for the plain data, encrypted data, and key pad respectively.}
  \label{fig:general_sch}
\end{figure}

For all these scenarios, chip 1 is assumed as the core component of the system, e.g., microprocessor, FPGA, etc. Thus, the cryptography capability assumption can be considered as valid for chip 1. If this capability is also supported by chip 2 (Figure 2(a)), EOP is fully applicable. This scenario requires no hardware modification on the original design. The conditionally applicable scenario (Figure 2(b)) occurs when chip 2 has no cryptography capability. In this scenario, EOP cannot be directly applied without the design modifications. One straightforward modification is replacing chip 2 with a crypto-enabled component (e.g., FPGA) if the original functionality of chip 2 is maintained.

If the modification mentioned above is not achievable, EOP can still be utilized to simplify board-level obfuscation \cite{guo2015investigation}. In the original obfuscation framework, an obfuscation block (chip 3 in Figure 2(c)) is inserted to permute the inter-chip communications. By encrypting the data paths between chip 1 and chip 3, the previously enforced design requirements (i.e., the BGA package and middle-lay routing) can be relaxed.    In this scenario, only the data flow between chip 1 and chip 3 is encrypted. Finally, the application scenario (a) can be combined with the board-level obfuscation as shown in Figure 2(d). Different from (c), all the data flows between chip 1, 2 and 3 are encrypted. EOP should be implemented independently for chip 1/chip 3 and chip 3/chip 2. In this scenario, chip 3 decrypts and re-encrypts the data with different keypads.

Figure \ref{fig:general_sch} shows the block diagram of EOP. This framework consists of four major modules: \textbf{encryption module}, \textbf{decryption module}, \textbf{control clock generation module}, and \textbf{control clock verification module}. The first two modules (referred as the crypto modules) encrypt and decrypt the messages from chip 1 and chip 2. The last two modules generate and verify the control clock to drive the first two modules. These modules are elaborated in the following sections. In Figure \ref{fig:general_sch}, the one-way communication is assumed between chip 1 and chip 2 (i.e., data are sent by chip 1 to chip 2). If the two-way communication is desired, the control clock generation and verification modules should be implemented in both chips. For simplicity, only the one-way communication scenario is discussed in this paper.

\subsection{Crypto Modules}
\label{sec:encryption modules}
In EOP, the encryption and decryption modules both consist of an $N$-bit Trivium cipher and an array of $N$ exclusive OR gates (XORs). The value $N$ equals to or greater than the number of data paths to be encrypted. The Trivium cipher generates a $N$-bit keypad vector $k$. \textcolor{black}{Besides the encrypting and decrypting tasks, these modules also take the responsibility of managing or synchronizing the key.}

For the sender side (chip 1 in Figure \ref{fig:general_sch}), each keypad bit, $k_n(t)$, is XORed with one data path ($p_n(t)$). As shown in the Figure \ref{fig:general_sch}(a), this operation generates a 1-bit cipher message, $e_n(t)$, for each data path. These cipher messages are then transmitted to the decryption module of chip 2. In the decryption module, the cipher messages are XORed with the keypad vector in a bit-wise manner. In normal operation, the Trivium ciphers in the encryption and decryption modules share the same keys as they employ the same keypads. To ensure that the keypad vectors are consistently synchronized on both the chips, the control clock modules are implemented. Details are elaborated in the following section.

\textcolor{black}{As mentioned in Section \ref{sec:trivium}, the Trivium cipher is initialized by loading an 80-bit key and an 80-bit initial vector. These values define the starting points of the stream cipher. Thus, the initial values should be synchronized between the encryption and decryption modules during each power-up. Additionally, these values should be changed for each power-up. This step is critical to guarantee that the correct data will be received at the receiver side. By changing the initial values during each power-up, the attacker cannot discover these values through the records of previous power-ups.}

\begin{figure}
  \centering
  \includegraphics[width=\linewidth]{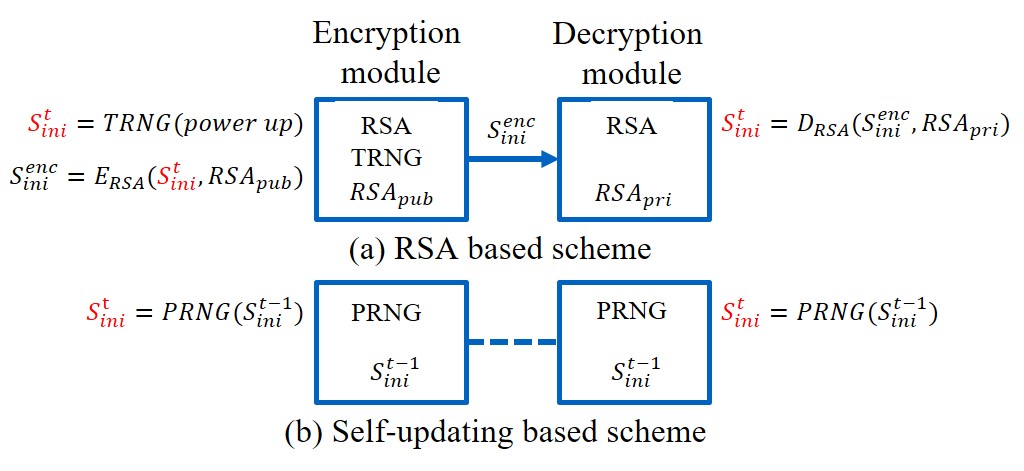}
  \caption{Seed initialization strategies: (a) share the seed with RSA; (b) seed self-updating.}
  \label{fig:seed ini}
\end{figure}

Two strategies can be implemented to accomplish this objective, as shown in Figure \ref{fig:seed ini}. For the RSA based scheme (\ref{fig:seed ini}(a)), the initial seed $S_{ini}^t$ is generated in the encryption module by a true random number generator (TRNG) \cite{fischer2002true}. Next, this seed is encrypted with the RSA public key ($RSA_{pub}$) and transmitted to the decryption module. In the decryption module, the initial seed can be recovered by the RSA private key ($RSA_{pri}$).  In this case, the RSA algorithm should be implemented in both the encryption and decryption modules. Moreover, a TRNG should be implemented in the encryption module.

For the self-updating based scheme shown in Figure \ref{fig:seed ini}(b), two pseudo-random number generators (PRNGs) with identical structure are implemented in the encryption and decryption module, respectively. The initial seed for both modules is updated by its previous value. When the firmware is loaded into the system by a trusted party, an arbitrary value is embedded as the first seed. This value will be utilized as the start state of the PRNG. At the $t^{th}$ power-up, the PRNG accepts the previous seed ($S_{ini}^{t-1}$) as the start state to generate the current seed ($S_{ini}^t$). To ensure the same seeds are synchronized in both encryption and decryption modules, the hash values of the current seeds may be computed. Next, the hash value in the encryption module is forwarded to the decryption model. By comparing its hash value and the received one, the decryption module verifies whether the same seed is generated.

\subsection{Control Clock Modules}
\label{sec:control clock}
The working mechanism of the \textbf{control clock generation} module is shown in Algorithm \ref{alg:clk gen}. The pseudocode is presented in the hardware description language (HDL) style. The module generates the control clock by taking the plain data paths as inputs. The \textbf{flipping event} time instances (i.e., $t_0$ and $t_1$ in Algorithm \ref{alg:clk gen}) are monitored and updated consecutively. The flipping event is defined as any value change of any data path, which is selected to be encrypted. For instance, assuming the number of data paths (i.e., $N$) is 8, a transit from $8'b00000000$ to $8'b00000001$ will be considered as a flipping event.

\begin{algorithm}[htpb]
\caption{Control clock generation}\label{alg:clk gen}
\algrenewcommand\algorithmicprocedure{\textbf{always}}
\begin{algorithmic}[1]
\State \textbf{Input}: plain data paths ($p_n(t)$)
\State \textbf{Output}: control clock ($clk$)
\State $t_0 \gets 0$ \Comment{\textit{initial flipping time instance}}
\Procedure{@}{any signal in $p_n(t)$ flips}
\State $t_0 \gets t_1$ \Comment{\textit{store previous flipping time instance}}
\State $t_1 \gets \text{current time instance}$
    \If{$t_1 - t_0 > thr$}
        \State produce \textbf{$clk$} \Comment{\textit{generate a control clock pulse}}
    \EndIf
\EndProcedure
\end{algorithmic}
\end{algorithm}

The control clock generation flow can be described as following. First, as shown in lines 5 to 6 in Algorithm \ref{alg:clk gen}, the previous flipping event time instance is stored, and the current value is updated. Next, the time interval between two contiguous flipping events (i.e., $t_0$ and $t_1$) are checked (line 7). Finally, a control clock pulse is produced when this time interval is longer than the threshold, $thr$. This pulse should have the minimum possible width. The steps as shown in lines 7 to 9 are utilized to prevent the potential glitches. These glitches may be induced by two flipping events which are too close to each other. The threshold for preventing the glitches can be assigned experimentally by the designer.

In general, the control clock generation module guarantees that a control clock pulse is produced when any data path changes its value. This control clock drives the stream cipher to produces the keypad vector and fetches new encrypted data. Thus, the encrypted data are generated slightly after the rising edge of the control clock pulse.

\begin{algorithm}[t]
\caption{Control clock verification}\label{alg:clk ver}
\algrenewcommand\algorithmicprocedure{\textbf{always}}
\begin{algorithmic}[1]
\State \textbf{Input}: encrypted data paths ($e_n(t)$)
\State \textbf{Input}: control clock ($clk$)
\State \textbf{Output}: verification status ($ver$)
\State $t_1^{data} \gets 0$ \Comment{\textit{initial flipping time instance}}
\Procedure{@}{$clk$ \textbf{or} any signal in $e_n(t)$ flips}

\If{$clk$ flips}
    \State $t^{clk} \gets \text{current time instance}$, then \textbf{Hold}
    \If{data path flipping event presents}
        \State $ver \gets \textbf{Safe}$
    \Else
        \State $ver \gets \textbf{Tampered}$
    \EndIf
\EndIf

\If{$e_n(t)$ flips}
    \State $t_0^{data} \gets t_1^{data}$
    \State $t_1^{data} \gets \text{current time instance}$
    \If{$t_0^{data} \le t^{clk}$ \textbf{and} $t_1^{data} \ge t^{clk}$}
        \State $ver \gets \textbf{Safe}$
    \Else
        \State $ver \gets \textbf{Tampered}$
    \EndIf
\EndIf

\EndProcedure
\end{algorithmic}
\end{algorithm}

Algorithm \ref{alg:clk ver} provides the function description of the \textbf{control clock verification} module. This module accepts the incoming encrypted data and control clock as its inputs. The output is the verification status which is either safe or tampered. In general, this module verifies whether the received control clock is unmodified after generated. To achieve this goal, the flipping events of the control clock and encrypted data path are monitored (line 5 in Algorithm \ref{alg:clk ver}). The verification status, $ver$, can be developed by comparing the recorded time instances. This comparison can be introduced as two situations concerning the type of flipping events detected. These situations are described as following.

\textbf{Control clock flipping event is detected (line 6 to 13)}: The latest control clock flipping event is recorded as $t^{clk}$ (line 7). Once this time instance is stored, the verification module holds briefly for a data path flipping event. This holding operation is implemented due to the short delay between the control clock and the encrypted data. If a data flipping event presents, the system can be considered as safe (line 8 to 10).

\begin{figure}
  \centering
  \includegraphics[width=\linewidth]{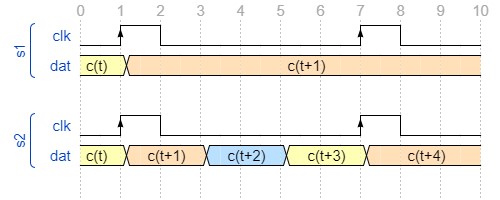}
  \caption{Example of control clock verification situations. Two violation scenarios are present as the following: (a) data missing and (b) control clock missing.}
  \label{fig:waveform vio}
\end{figure}

\textbf{Data path flipping event is detected (line 14 to 22)}: As shown in line 15 to 16, the time instances of the latest and its contiguous encrypted data flipping events are recorded as $t_0^{data}$ and $t_1^{data}$. Besides these two time instances, the latest control clock flipping events is also recorded as $t^{clk}$ (line 7). An example can be found in Figure \ref{fig:waveform vio}. This example shows a safe verification status (i.e., $ver \leftarrow Safe$). According to the waveform group s2 in this figure, $t_0^{data}$ is recorded at time instance 5. $t_1^{data}$ is recorded slightly behind time instance 7. The $t^{clk}$ is recorded at time instance 7. The verification module compares the time instances of the current and previous data path flipping events with the control clock flipping event. Only if clock flipping event is observed in between of two adjacent data path flipping events, it implies that the system is safe (line 17 to 18).

Two violation scenarios are provided in Figure \ref{fig:waveform vio}. The time instances from $t=0$ to $t=10$ are shown at the top of the figure. For the waveform group s1, a control clock flipping event is detected at $t=7$. However, no data path flipping event is detected after this control clock pulse. Thus, the tampering activity should be considered as happening at $t=7$. This tampering activity can be either an external hold of the encrypted data or a fake control clock pulse. For the waveform group s2 in Figure \ref{fig:waveform vio}, an encrypted data flipping event is recognized at $t=3$. Its previous flipping event is recorded at $t=1$. However, no control clock flipping event is observed between the time instances 1 and 3. Thus, tampering on either the encrypted data path or the control clock should be considered as happening at $t=3$.

\section{Attack Analysis}
\label{sec:attacks analysis}
In the section, the capability of EOP against various attacks is discussed. These discussions are organized by different application scenarios shown in Figure \ref{fig:application_scenario}. These scenarios are roughly classified into two categories: (i) partial encryption (scenario (c)) and (ii) full encryption (scenario (a) and (d)). For the partial encryption scenario, the linear stream cipher is also considered as the keystream generation option for EOP. By analyzing this option, we illustrate that the non-linear stream cipher is crucial in this scenario.

\subsection{Partial encryption}
\label{sec:partial encryption}

\begin{figure*}[t]
  \centering
  \includegraphics[width=\linewidth]{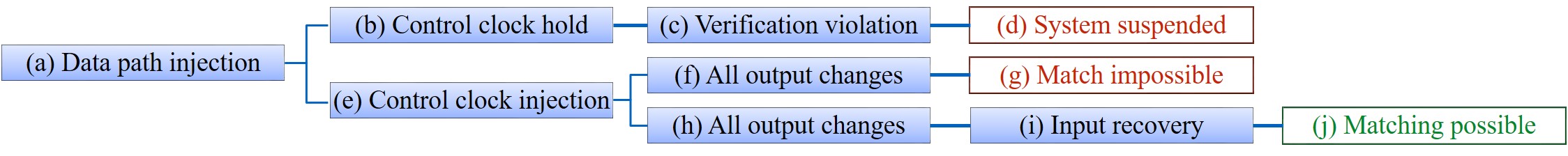}
  \caption{An example of the attack on EOP when the linear stream cipher is engaged.}
  \label{fig:linear cipher attack}
\end{figure*}

\begin{figure}[t]
  \centering
  \includegraphics[width=0.9\linewidth]{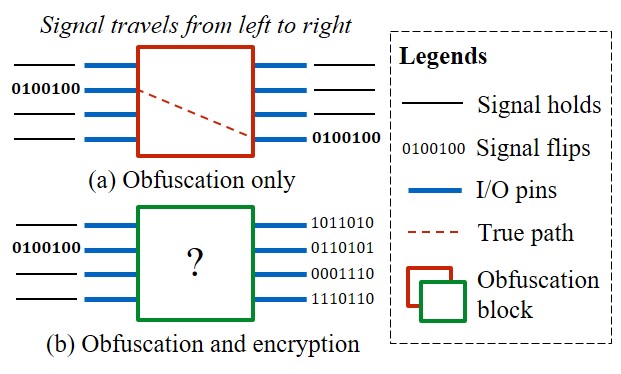}
  \caption{Tampering attacks on the obfuscation block: (a) only the permutation is implemented, and (b) EOP is combined with the permutation.}
  \label{fig:obfus vs obfus enc}
\end{figure}

The application scenario (c) in Figure \ref{fig:application_scenario}(c) is referred to as partial encryption since only the data flow between chip 1 and chip 3 is encrypted. The data flow between chip 3 and chip 2 is unprotected due to the crypto capability limitation of chip 3. Thus, only the reverse engineering attack is considered in this scenario. As discussed in Section \ref{sec:background}, the linear stream cipher is not secure enough due to the limited number internal states. To justify that the utilization of non-linear stream cipher is necessary, the reverse engineering attack is investigated against both types of the stream ciphers.

In this application scenario, the obfuscation block (i.e., chip 3) consists of the permutation block, decryption model, and the control clock verification model. It decrypts the received data by XORing them with the keypad. These decrypted data are then sent to chip 2 directly without re-encryption.

Since the obfuscation is enforced to prevent reverse engineering, the connections between chip 1 and chip 2 are permuted by the obfuscation block. Thus, the attacker's goal is to discover the correct input-output relationships of the obfuscation block. To achieve this goal, three attacking scenarios are present in Figure \ref{fig:linear cipher attack}.

The most straightforward attack flow is shown as $(a)\rightarrow(b)\rightarrow(c)\rightarrow(d)$ in Figure \ref{fig:linear cipher attack}. The attacker injects a periodic signal into one input of chip 3 and monitors the outputs for a match. This attacking scheme is illustrated in Figure \ref{fig:obfus vs obfus enc}(a). The matched signal path is shown as the red dashed line. However, this attack cannot be applied directly since solely changing the data path leads to a violation of the control clock verification module (as noted in Algorithm \ref{alg:clk ver}, line 12). To avoid such violation, a more advanced attack flow is $(a)\rightarrow(e)\rightarrow(f)\rightarrow(g)$, in which a pulse is also injected to the control clock while a data path flipping event presents. If the decryption module receives a control clock pulse, it updates the keypad. This keypad is constructed by combining the keystream bits from multiple linear stream ciphers in parallel. Thus, all the outputs will change their values at the same time when a control clock is received. This attacking scheme is presented in Figure \ref{fig:obfus vs obfus enc}(b).
Since the attacker injects the periodic signal into one input (the rest of the inputs are hold low), only one output is expected to flip for a successful match. However, the control clock pulses force all the outputs flip at the same frequency. In this case, it is impossible to achieve this match. Thus, the attack is defeated.

In order to achieve the matching while passing the control clock verification, the attack flow $(a)\rightarrow(c)\rightarrow(h)\rightarrow(i)\rightarrow(j)$ can be applied. The core step of this flow is recovering the inputs from the monitored outputs (step (i) in Figure \ref{fig:linear cipher attack}). To accomplish this recovery, the attacker needs to apply XOR on the output with the current keypad. Note that the keypad is generated by the linear stream cipher such as LFSR and its structure is assumed as public.  Thus, the next keypad is predictable when the current state is obtained. Since the attacker has access to both the plaintext and ciphertext, he can obtain an arbitrary length of the keystream bits. For the LFSR, the keystream bits are equal to the state bits. Thus, the current state can be obtained by exhaustive search (i.e., brute force), and the rest of the keystream bits can be generated. Finally, the keypad, which is constructed by the keystream bits, can be predicted. With this predicted keypad, the attacker can recover the inputs of obfuscation block from the outputs and perform the matching.

To eliminate this attack, the stream cipher must perform the non-linear feedback function with a sufficient number of internal states. For instance, the Trivium cipher consists of a 288-bit internal state register which provides more than $4.97e+86$ internal states. Concerning the current computing power, this number is sufficient for preventing brute force attacks.

\subsection{Full encryption}
\label{sec:full encryption}
Both the application scenarios (a) and (d) in Figure \ref{fig:application_scenario} are referred to as full encryption, which means that all the critical data flows are encrypted. For both scenarios, the passive and active tampering attacks are analyzed. For the scenario (d), the reverse engineering attack is also considered besides the tampering attacks. Being introduced in Section \ref{sec:intro}, the load station in the Cisco router is the device to achieve the passive tampering. This device is implanted by the attacker and eavesdrops the user data. These user data are sent to the attacker by the load station. The Xbox mod-chip can be viewed as an example of the active tampering. Unlike the load station, the Xbox mod-chip not only monitors the user data but also modifies these data. These modified data create fake messages to the system (e.g., mislead the copyright check module). In most of the cases, the active tampering consists of the passive tampering and data modification. Thus, once the active tampering is prevented, the passive tampering is eliminated.

Active tampering manipulates the inter-chip communications by injecting designated data such as malicious signals to bypass the authentication process. By implementing EOP, the attacker cannot inject meaningful data without knowing the keypad. In other words, the decrypting equation of the data received by the target chip behaves like a black box for the attacker. For the passive tampering, the attacker only has access to the encrypted data flow. Similar to the active tampering case, the attacker cannot obtain the plaintext without the knowledge of the keypad. As stated in the previous section, the keypad cannot be predicted when the non-linear stream cipher is employed.

For the reverse engineering attack, the full encryption scenarios provide even stronger protection compared with the partial encryption scenario. Different from application scenario (c), the application scenario (d) consists of two encryption-decrypting processes (i.e., chip 1 to chip 3 and chip 3 to chip 2). In this situation, the data flow is encrypted by chip 1 and sent to the obfuscation block (chip 3). Next, this obfuscation block decrypts the received data flow and re-encrypts it. Eventually, the re-encrypted data flow is sent to chip 2 for the final decryption. These two encryption-decrypting processes are initialized by different key and IVs (initial vectors). Additionally, the attacker only has access to the ciphertext of both processes. Thus, it is impossible to predict the keypads and perform the matching.

In summary, the attacks described in Section \ref{sec:intro} are fully addressed by EOP. The tampering attack, such as the modchip and load station implanting, are eliminated by the full encryption scenarios. The reverse engineering attacks are defeated by combining with the board-level obfuscation \cite{guo2015investigation} with EOP. By taking advantages of EOP, the implementation of the obfuscation is significantly simplified from a PCB design perspective.

\section{Experimental Results}
\label{sec:resutls}
In this section, the following experiments are conducted to validate the performance of proposed framework. To verify that the encrypted data can be decrypted correctly, these data are generated under different conditions. This test confirms that EOP maintains the correct system functionality when no tampering attacks are occurring. Besides the normal operation test, we apply the active tampering attack on the data path and the control clock to demonstrate the tampering detection sensitivity. These experimental results are collected from both simulations and the hardware implementations. Before providing these results, the experimental platform setups are introduced.

\subsection{Platform Setups}
\label{sec:platform}
\begin{figure}[t]
  \centering
  \includegraphics[width=0.8\linewidth]{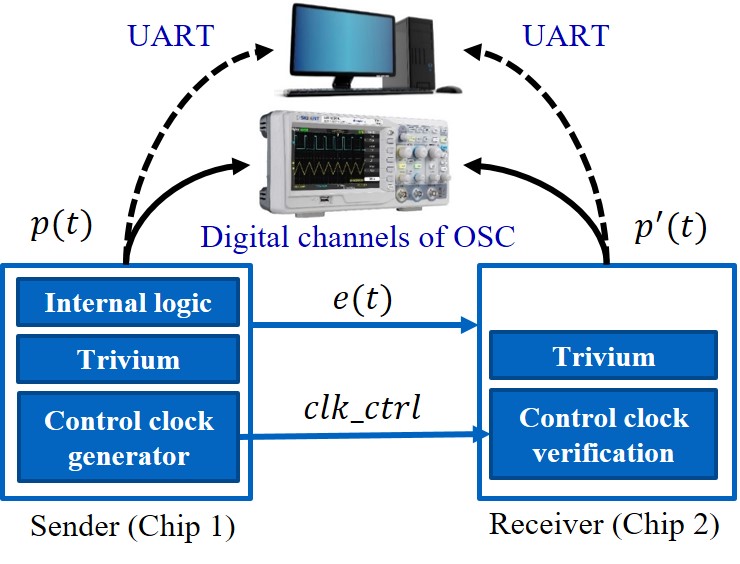}
  \caption{Schematic of the testing environment. The original plain data ($p$) and decrypted data ($p^\prime$) are captured by the digital oscilloscope and collected by the PC.}
  \label{fig:test setup}
\end{figure}

Figure \ref{fig:test setup} shows a schematic setup of the testing platform. In our experiment, the application scenario (a) in Figure \ref{fig:application_scenario} is implemented. In this setup, the sender (chip 1) is assumed to send critical data to the receiver (chip 2). This sender consists of three modules:
\begin{itemize}
  \item The \textbf{internal logic} module generates the random signals with a width of 8 bits. These signals are utilized to mimic the plaintext data ($p(t)$) which is generated by the sender.
  \item The \textbf{Trivium} module takes the following responsibilities: updating the keypads and performing the XOR operation between the plaintext data and the keypad to produce the encrypted data ($e(t)$).
  \item The \textbf{control clock generate} module accomplishes the function which is described in \ref{sec:control clock}. This module takes the signal from the internal logic module and produces the control clock ($clk\_ctrl$).
\end{itemize}

\begin{figure}[b]
  \centering
  \includegraphics[width=\linewidth]{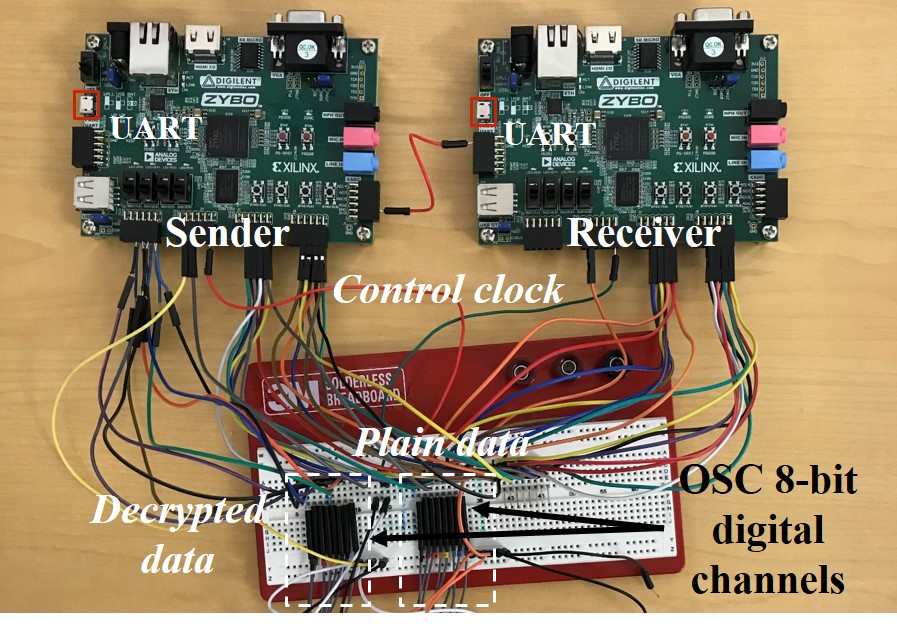}
  \caption{Hardware setups of the full applicable scenario (Figure \ref{fig:application_scenario}(a))}
  \label{fig:hardware}
\end{figure}

For the sender, the encrypted data ($e(t)$) and the control clock ($clk\_ctrl$) will be sent to the receiver. At the same time, the plaintext data ($p(t)$) are transmitted to the host PC through UART for further analyses. $p(t)$ is also monitored by the digital channels of the oscilloscope (OSC) for the debug purpose.

Different from the sender, the internal logic module is not implemented in the receiver (chip 2). For the receiver, the Trivium module performs the similar functionality but produces the decrypted plaintext data ($p^\prime(t)$). $p^\prime(t)$ is sent to the host PC and monitored by the OSC for comparison. The received control clock is also verified by the receiver. A counter is implemented in the control clock verification module to record the number of tampering induced violations.

The actual hardware implementation described above schematic is shown in Figure \ref{fig:hardware}. The Xilinx ZYBO development boards with the Zynq SoC are utilized to implement the sender and receiver. This SoC consists of a single-core ARM Cortex-A9 processor and a 28nm Artix-7 based programmable logic (PL). The modules shown in Figure \ref{fig:test setup} are implemented in the PL, while the processor configures and collects the signals from the PL. It serves as a bridge between the PL and the host PC. \textcolor{black}{This setup is straightforward to be applied on the commercial PCBs with COTS components.}

A breadboard is used to connect the sender and receiver through jumper wires. The digital channels of the OSC are connected with this breadboard to monitor the plaintext data and decrypted data. These data are also sent to the host PC through the onboard UART.

\subsection{Validation Results}
\label{sec:validation}
For validating the normal operation of EOP, various plaintext data are generated by the internal logic of the sender. These signals are generated by multiple LFSRs with the following characteristics: (i) Uniform frequency: all the 8 data paths update their values at the same frequency. These values are determined by the outputs of the 8 LFSRs. A single clock drives these LFSRs for the next state. (ii) Random frequency: different from the former case, the LFSRs are driven by eight clocks with random frequencies. The frequencies of these LFSR clocks are controlled by the SoC and range from $5MHz$ to $200MHz$ in a step size of $10MHz$.

Since different signal generation modes and frequencies are utilized, we apply the following three phases to conduct the experiment: initialization, execution, and pause. During the \textbf{initialization} phase, PL waits for the initial configurations (i.e., frequencies) from the processor. This is the default phase and the test system enters this phase upon startup. The \textbf{execution} phase sends a global start signal to the PL. All the modules start their functionalities based on the initial configurations. For the \textbf{pause} phase, all modules halt their current states and wait for further phase selection signal. During this phase, the counter in the receiver sends the number of violations to the PC.

The host PC collects both p(t)$ and p\prime(t)$ once the communication is established between the sender and the receiver (i.e., execution phase is set). $p(t)$ and $p\prime(t)$ are compared instantly when their values are changed. As the nominal operation validation, no tampering is applied, and the validation target is the decryption correctness. The full encryption-decryption procedure is constantly executed for one hour under all frequency corners. The aforementioned comparison between $p(t)$ and $p\prime(t)$ are made. The number of mismatches is reported to the PC at the end of each evaluation. The experimental results demonstrate that the receiver can consistently decrypt the correct data and no decryption errors are found during all the experiments.

\begin{figure*}[t]
  \centering
  \includegraphics[width=\linewidth]{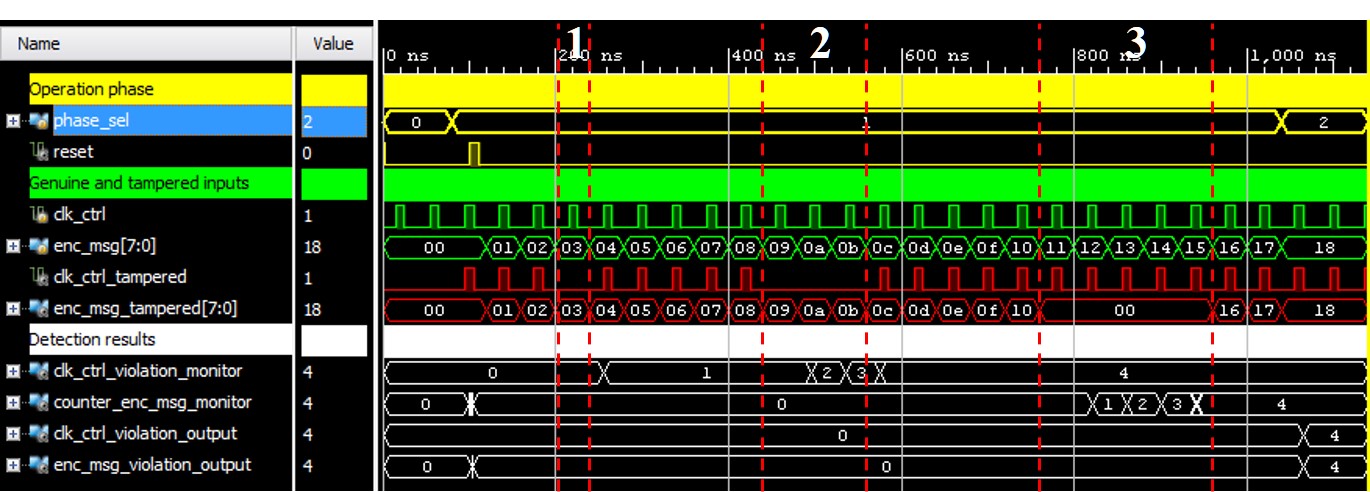}
  \caption{Simulation results for the active tampering.}
  \label{fig:sim result}
\end{figure*}

Aside from the nominal operation validation, we also conduct the active tampering attacks on EOP. Both the data paths and the control clock are manipulated in our experiment. Prior to presenting the silicon results, the post-implementation simulation results are shown in Figure \ref{fig:sim result}. This post-implementation simulation is accomplished by the Xilinx Vivado ISim. Since it considers the physical properties of the design, the timing is close to the silicon results. In this figure, the yellow signals indicate the current phase (i.e., initialization, execution, and pause phase), while the green signals refer to the generic signals, the red ones refer to the tampered signals, and the white signals show the number of violations.

\begin{table}[t]
\centering
\caption{Violation counts under different tampering durations.}
\label{tab:violation counts}
\renewcommand{\arraystretch}{1.3}
\begin{tabular}{|c|c|c|c|c|c|}
\hline
\multicolumn{2}{|c|}{Tampering durations (5ns base)} & $1 \times$ & $5 \times$ & $10 \times$ & $100 \times$ \\ \hline
\multirow{2}{*}{Average violations} & Control clock & 1.02 & 5.08 & 10.05 & 100.01 \\ \cline{2-6}
 & Data path & 1.05 & 5.10 & 10.10 & 100.07 \\ \hline
\end{tabular}
\end{table}

\begin{table}[t]
\centering
\caption{Technical comparison}
\label{tab:technical comparison}
\renewcommand{\arraystretch}{1.3}
\begin{tabular}{ccccc}
\hline
\multicolumn{1}{|c|}{Metrics} & \multicolumn{1}{c|}{\begin{tabular}[c]{@{}c@{}}Security\\ enclosure \cite{isaacs2013tamper}\end{tabular}} & \multicolumn{1}{c|}{\begin{tabular}[c]{@{}c@{}}Obfuscation \\ \cite{guo2015investigation}\end{tabular}} & \multicolumn{1}{c|}{\begin{tabular}[c]{@{}c@{}}MPA \\ \cite{guo2017mpa}\end{tabular}} & \multicolumn{1}{c|}{\begin{tabular}[c]{@{}c@{}}EOP\end{tabular}} \\ \hline
\multicolumn{1}{|c|}{\begin{tabular}[c]{@{}c@{}}Crypto\\ models\end{tabular}} & \multicolumn{1}{c|}{No} & \multicolumn{1}{c|}{No} & \multicolumn{1}{c|}{No} & \multicolumn{1}{c|}{Yes} \\ \hline
\multicolumn{1}{|c|}{\# of RNGs} & \multicolumn{1}{c|}{0} & \multicolumn{1}{c|}{0} & \multicolumn{1}{c|}{0} & \multicolumn{1}{c|}{1} \\ \hline
\multicolumn{1}{|c|}{\begin{tabular}[c]{@{}c@{}}Implement\\ overhead\end{tabular}} & \multicolumn{1}{c|}{High} & \multicolumn{1}{c|}{Medium} & \multicolumn{1}{c|}{Medium} & \multicolumn{1}{c|}{Low} \\ \hline
\multicolumn{1}{|c|}{\begin{tabular}[c]{@{}c@{}}Other\\ requirements\end{tabular}} & \multicolumn{1}{c|}{\begin{tabular}[c]{@{}c@{}}Limited\\ applications\end{tabular}} & \multicolumn{1}{c|}{\begin{tabular}[c]{@{}c@{}}Package \\ and routing\end{tabular}} & \multicolumn{1}{c|}{\begin{tabular}[c]{@{}c@{}}HIGH-Z\\ JTAG\end{tabular}} & \multicolumn{1}{c|}{None} \\ \hline
\multicolumn{1}{|c|}{\begin{tabular}[c]{@{}c@{}}SR \\ coverage\end{tabular}} & \multicolumn{1}{c|}{\begin{tabular}[c]{@{}c@{}}SR-1, \\ SR-2, SR-3\end{tabular}} & \multicolumn{1}{c|}{SR-4} & \multicolumn{1}{c|}{\begin{tabular}[c]{@{}c@{}}SR-1, \\ SR-2\end{tabular}} & \multicolumn{1}{c|}{ALL SR*} \\ \hline
\multicolumn{5}{l}{\textit{\begin{tabular}[c]{@{}l@{}}* See the text for the detailed SR coverage considering \\ different application scenarios\end{tabular}}}
\end{tabular}
\end{table}

During the execution phase ($phase\_sel=1$), three tampering attacks are applied. These attacks are separated by the red vertical dashed lines and numbered as 1, 2, and 3 in this figure. Among all the tampering attacks, 1 and 2 indicate the control clock tampering attack. Attack 3 refers to the data path tampering attack. All the attacks attempt to force the signal to ground. In total, four control clock violations are applied (one during attack 1 and three during attack 2). All the four data path tampering violations are contributed by the attack 3. The last two signals in Figure \ref{fig:sim result} confirm the functionality of the control clock verification block.

Besides the simulations results, these tampering attacks are conducted on the actual hardware. A function generator and tristate buffers are utilized to tamper the connections between the sender and receiver actively. The function generator controls the tristate buffers to ground these connections for different durations. The basic unit of these durations is the shortest period for the sender to update the plain data (i.e., 200Mhz/5ns). The testing conditions and results are presented in Table \ref{tab:violation counts}.

The first row in Table \ref{tab:violation counts} indicates the tampering durations which are expressed as how many times of the smallest data path updating period (i.e., 5ns). Theoretically, one violation should be counted once when the tampering period is 5ns. However, since this tampering signal is not aligned with control clock and may last slightly longer than designed, more than one violation might be recorded. This situation is rare, and only appears when the tampering occurs right after a rising edge of the control clock. Since its duration may be longer than 5ns, it may also affect the next control clock rising edge.

Table \ref{tab:technical comparison} compares the previous and proposed techniques which are designated to meet the security requirements, SR-1 to SR-4. Among all the techniques, only EOP demands the cryptography modules and the random number generator (RNG). These components are utilized to initial  and accomplish the encryption process. The security enclosure \cite{isaacs2013tamper} introduces the highest implementation overhead due to the demands of the hard metal case and sensors. Since only one additional chip (e.g., CPLD/FPGA) is installed, the techniques (obfuscation \cite{guo2015investigation} and MPA \cite{guo2017mpa}) are considered as medium overheads. EOP presents the lowest overheads considering no extra components might be introduced. This claim is true besides achieving the anti-reverse engineering requirement (SR-4).

Considering the implementation restrictions, the security enclosure \cite{isaacs2013tamper} can be only implemented in a limited number of applications. These applications should not experience frequent movements and temperature variations. Thus, security enclosure is not suitable for many consumer electronics. The obfuscation based approach \cite{guo2015investigation} requires specified packages (e.g., BGA) and dedicated routing schemes (i.e., routing in the middle layers). MPA \cite{guo2017mpa} configures the pins' JTAG infrastructure into the HIGH-Z mode. However, this mode is an optional mode of the JTAG standard and may not be universally available.. For EOP, there are no significant constraints for the PCB.

For the SR coverage (detailed SR definitions in Section \ref{sec:sr}), since the security enclosure targets the protection against tampering, it fails to satisfy SR-4 (anti-reverse engineering). The same situation is applied for the MPA \cite{guo2017mpa} which is only designed against tampering. Additionally, the run-time requirement (SR-3) is not applicable on MPA since it requires a detecting phase. During this phase, the system cannot perform its normal function. For the obfuscation based approach \cite{guo2015investigation}, the BGA package and middle-layer routing only protect part of the critical paths from tampering. The paths between the obfuscation block and the non-programmable components cannot be hidden because few non-programmable components have BGA package. Among all the techniques in Table \ref{tab:technical comparison}, only EOP (application scenario (d)) thoroughly covers all the SRs.

\section{Conclusion}
\label{sec:conclusion}
We proposed a framework named EOP to mitigate multiple practical attacks on a PCB, such as tampering, hardware Trojan, and reverse engineering. EOP encrypts the inter-chip communications by keypads. These keypads are generated by the non-linear stream cipher and updated with the control clock. The switch activities of this control clock and data paths are correlated. This correlation is verified by a dedicated module against the tampering attacks. EOP can be utilized to combine with the board-level obfuscation technique to provide the most comprehensive solution for the PCB-level assurance. For validation, EOP is implemented in Zynq SoCs. Various operation frequencies and tampering detection performance are verified. The results confirm the detection accuracy of EOP.



\bibliographystyle{IEEEtran}
\bibliography{real_time_encryption}
%



\end{document}